\documentclass[journal=nalefd,manuscript=article,layout=twocolumn]{achemso}

\usepackage[version=3]{mhchem} 
\usepackage{latexsym} \usepackage{amsmath} \usepackage{dcolumn}
\usepackage{bm} \usepackage{graphicx} \usepackage{epsfig}
\usepackage{float}

\usepackage[]{graphicx}
\usepackage[]{wrapfig}
\usepackage[]{color}

\author{Towfiq Ahmed}
\email{atowfiq@u.washington.edu}
\affiliation
[Theoretical Division, Los Alamos National Laboratory]{Theoretical Division,
Los Alamos National Laboratory, Los
Alamos, New Mexico 87545}
\author{Jason T. Haraldsen}
\affiliation
[
Department of Physics and Astronomy, James Madison University]{Department of Physics and Astronomy,
James Madison University, 
Harrisonburg, VA 22807}
\author{John J. Rehr}
\affiliation[Department of Physics, University of Washington]{Department of
Physics, University of Washington, Seattle Washington 98195}
\author{Massimiliano Di Ventra}
\affiliation
[Department of Physics, University of California, San Diego]{Department of
Physics, University of California, San Diego, California 92093}
\author{Ivan Schuller}
\affiliation
[Department of Physics, University of California, San Diego]{Department of
Physics, University of California, San Diego, California 92093}
\author{Alexander V. Balatsky}
\email{avb@lanl.gov}
\affiliation
[Nordita
Roslagstullsbacken 23, 106 91 Stockholm
Sweden]
{Nordic Institute for Theoretical Physics, KTH Royal Institute of Technology and Stockholm University,  Stockholm, Sweden}
\altaffiliation
{Theoretical Division,
Los Alamos National Laboratory, Los
Alamos, New Mexico 87545}
\altaffiliation
{Center for Integrated Nanotechnologies, Los Alamos
National Laboratory, Los Alamos,
New Mexico 87545}

\title[\texttt{achemso}]
{Correlation dynamics and enhanced signals for serial DNA sequencing}


\begin{document}

\begin{abstract}
Nanopore based sequencing  has demonstrated significant potential for the
development of fast, accurate, and cost-efficient fingerprinting techniques for
next generation molecular detection and sequencing. We propose a specific
multi-layered graphene-based nanopore device architecture for the
recognition of
single DNA bases. Molecular detection and analysis can be accomplished
through the detection of transverse currents as the molecule or DNA base
translocates through the nanopore. To increase the overall signal-to-noise
ratio and the accuracy, we implement a new ''multi-point cross-correlation''
technique for identification of DNA bases or other molecules on the
molecular level. we demonstrate that the cross-correlations between each
nanopore will greatly enhance the transverse current signal for each molecule.
We implement first-principles transport calculations for DNA bases surveyed
across a multi-layered graphene nanopore system to illustrate the advantages
of proposed geometry.  A time-series analysis of the cross-correlation
functions illustrates the potential of this method for enhancing the
signal-to-noise ratio. This work constitutes a significant step forward
in facilitating fingerprinting of single biomolecules using solid state
technology.
\end{abstract}

\section{Introduction}
\begin{figure}[!ht]
\includegraphics[width=3.5in]{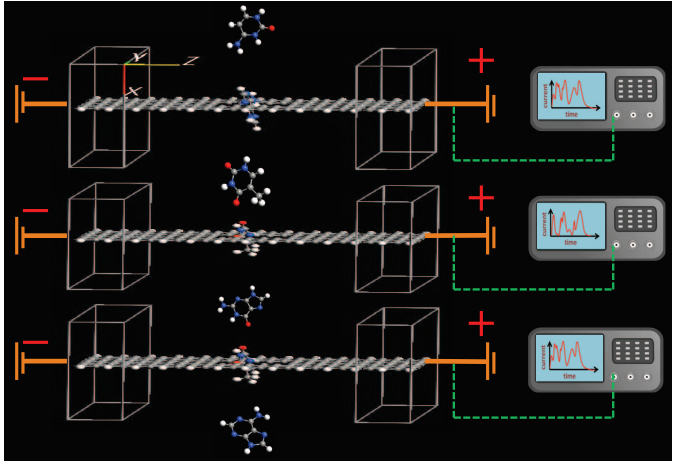}
\caption{\small {\bf {a}} Illustration of the three layer graphene based nanopore as a possible multilayered sequencing device.
{\bf {b}} Schematic of transmission currents through two graphene
layers where isolated DNA bases pass through the nanopores. The current vs. time spectra are recorded for each layer independently.  A cross correlation
between the current data from multipores reveals useful information by
increasing signal to noise ratio as described in the text; {\bf {c}} Hydrogen
capped graphene nanoribbons and the DNA
bases inside the pore. Here, only the flat orientation of the DNA bases
are shown. The other angular orientations are shown in the supplementary section.are shown. The other angular orientations are shown in the supplementary section.}
\label{f1}
\end{figure}

 With applications ranging from explosives and drug detection to DNA sequencing and biomolecular identification, the ability to detect specific molecules and/or molecular series presents many challenges for scientists. With a specific need for timely and accurate measurements and evaluation, it is essential that researchers investigate both the manner of detection as well as explore new and improved computational methods for analysis to keep up with the growing pace of the individual fields.

The field of DNA sequencing is rapidly evolving due to increasing support and technology. As this occurs, sequencing techniques are challenged by the need for
a rapid increase of accuracy, speed, and resolution for smaller amounts of
material~\cite{xprize}. Nanopore-based sequencing~\cite{Zwolak2008,Branton2008}
and serial methods~\cite{towfiq_dna,Kilina2007} provide promising alternatives to the well established Sanger
method~\cite{sanger}, particularly for
identifying single DNA bases using transverse conductance~\cite{ventra_2005,ventra_2006}. Such an approach
relies on the ability to resolve the electronic fingerprints of DNA
one relevant unit
at a time (`serial') as DNA translocates through a nanochannel. It has been established that
experimental methods are capable of achieving single-base resolution, which has prompted investigations into the local electrical
properties of single DNA bases~\cite{tanaka,Yarotski2009}.
Concurrently, the theoretical underpinnings of this approach have been continuously developing~\cite{towfiq_dna,Kilina2011,Kilina2007,ventra_2005,ventra_2006}.

The
single-molecule sensitivity of nanopore sequencing has been recently demonstrated by
Kawai {\it {et al.}}~\cite{kawai} and Lindsay{\it {et al.}}~\cite{Chang2010}. The sequence of DNA/RNA oligomers and microRNA by tunneling has also been
demonstrated~\cite{kawai2012}.
Despite such high-quality
experimental methods,
the most pressing challenge in serial sequencing lies in overcoming effects of noise that lead to a small signal to noise (S/N) ratio
in the measured current $I$. The signal fluctuations generally originate
from thermal
agitation and bond formation between base and nanopore/electrode walls or interactions with a substrate.
In an effort to avoid these limitations,
we propose the sequential measurement of transverse current cross-correlations, as obtained from multiple pairs of electrodes. The experimental set up for such a nanopore arrangement is schematically shown in \ref{f1}.
To be specific, we focus on graphene as the porous material, because it is atomically thick and exhibits extraordinary thermal
and electronic
properties.
\begin{figure*}[htpb]
    \begin{minipage}[!t]{0.70\linewidth}
   \epsfig{file=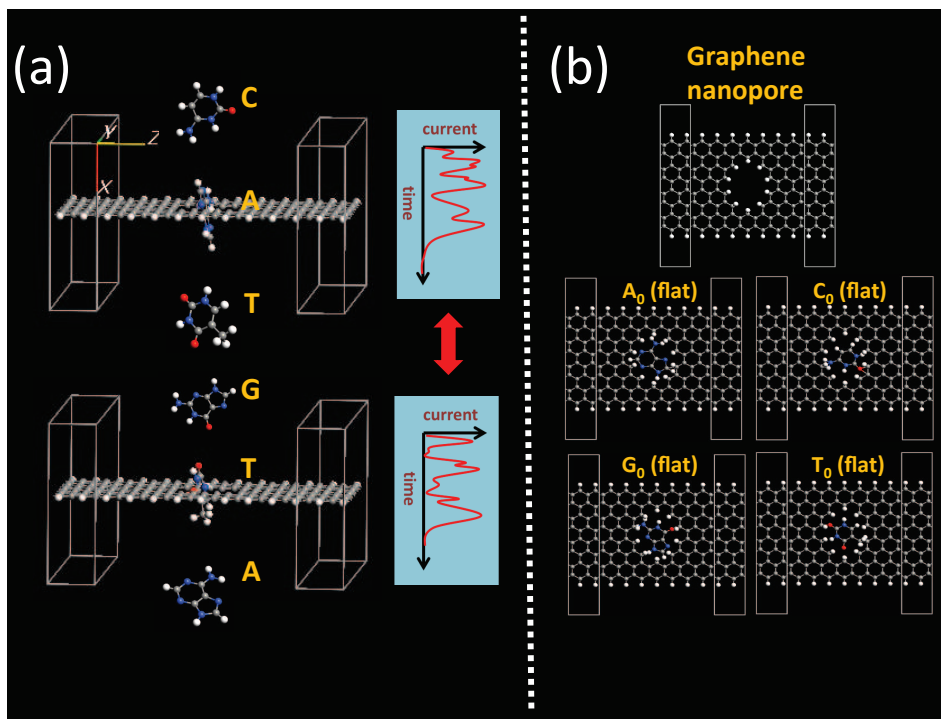, width=\linewidth}
    \end{minipage}\hfill
    \begin{minipage}[!t]{0.28\linewidth}
\caption{\small {\bf {a}} Schematic of transmission currents through two graphene
layers where isolated DNA bases pass through the nanopores. The current vs. time spectra are recorded for each layer independently.  A cross correlation
between the current data from multipores reveals useful information by
increasing signal to noise ratio as described in the text; {\bf {b}} Hydrogen
capped graphene nanoribbons and the DNA
bases inside the pore. Here, only the flat orientation of the DNA bases
are shown. The other angular orientations are shown in the supplementary section.}
\label{f2}
\end{minipage}
\end{figure*}
Besides these geometric advantages and good conductivity, graphene
also
possesses high tensile strength and can endure a high transmembrane pressure environment~\cite{graphene_mechanical}.
Consequently, graphene has been proposed as an effective substrate and conducting medium
for nanopore sequencing by numerous
groups~\cite{branton,merchant,schneider,prezhdo,tanaka,scheicher}.
We emphasize, however, that the method for nanopore sequencing may be useful in any other method in which serial measurements (e.g., time series)
are made to ascertain individual properties (resistivity here) of the bases.

Although this challenge is much more severe for protein based or solid state
nanopores, the nature of an
atomically thick graphene nanopore wall cannot completely rule out the
$\pi-\pi$ stacking between carbon and DNA bases. In addition, vibration and
other electronic fluctuations present in the graphene membrane can significantly mask
the
conductance signals, making it difficult to differentiate the individual
DNA bases.

Previous theoretical~\cite{Kilina2007,Kilina2011} and experimental~\cite{tanaka} studies of the
interactions between DNA bases and graphene derivatives have revealed the
local electronic
structure of single bases.
The experimental realization of a single layer graphene-based nanopore
device is made possible by combining
several state of the art techniques e.g., mechanical exfoliation
from graphite on SiO$_2$ substrate.
Transverse tunneling current(conductance)
measurements, as the single strand (ss)DNA translocates through a monolayer graphene nanopore, were
previously reported by Schneider {\it et al.}~\cite{schneider}.
AFM studies~\cite{Yarotski2009}
and theoretical simulations
of scanning tunneling spectroscopy (STS)~\cite{towfiq_dna} support the
identification of
electronic features with varying spatial extent and intensity near the
HOMO-LUMO band.

To make nanopore sequencing and detection a viable method for determining translocating molecules, one must overcome this the noise to signal problem.
Therefore, we propose a multilayered graphene device in which the transverse conductance is measured
through each nanopore independently, as a series of DNA bases or other molecules translocates through them (see ~\ref{f1}).
As molecules translocate, they create a time dependent sequence of
translocation currents through each of the layers. One then monitors the translocation
currents at
different pores and acquires a record of sequential current of the same base as
it arrives and moves through the individual pores (shown in ~\ref{f2}). The time series of the cross correlation currents can then
be used to reduce the uncorrelated, independent noise source, and hence enhance the signal to noise ratio and
improve the differentiation between bases. While our device is being discussed under the idea of DNA sequencing, the general method and device setup can be used for any molecule small enough to fit through a nanopore. While we are focusing on the area of DNA sequencing and biomolecules, this cross-correlation method for data analysis of the transverse currents can be utilized for the analysis of any molecular series given the proper understanding of the molecules electronic properties.

\begin{figure*}[htpb]
    \begin{minipage}[!t]{0.70\linewidth}
   \epsfig{file=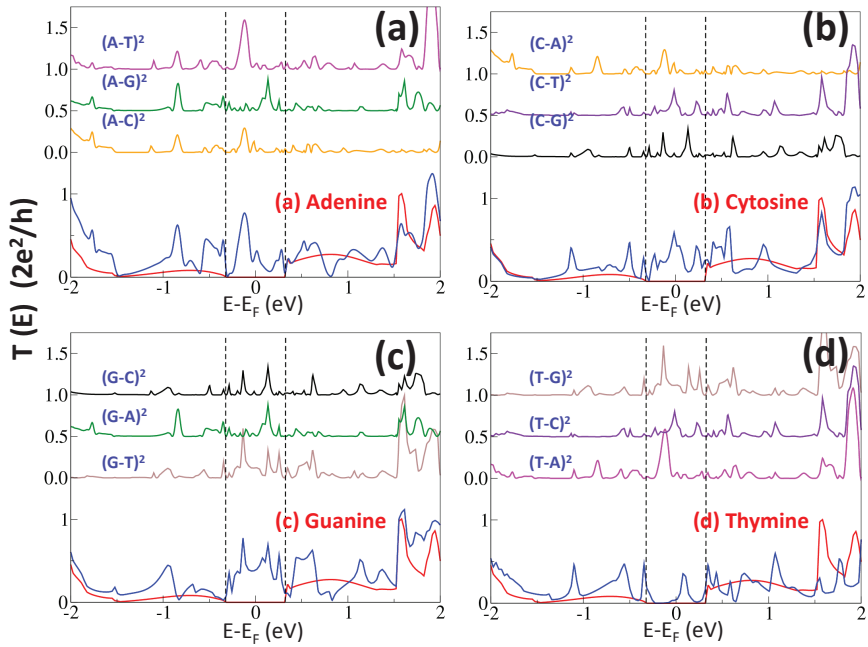, width=\linewidth}
    \end{minipage}\hfill
    \begin{minipage}[!t]{0.28\linewidth}
\caption{\small {Configuration averaged transmission coefficients (solid blue lines) for
{\bf{(a)}} Adenine,
{\bf{(b)}} Cytosine,
{\bf{(c)}} Guanine, and
{\bf{(d)}} Thymine.
The solid red line is T(E) for pure graphene with nanopore for comparison.
The vertical dashed lines are at -0.35 eV and +0.35 eV which are the
E$_F$ of the left and right electrodes respectively. The top three curves
in each panel are the difference-square curves between the average T(E) for
each base.
The fermi energy of the central region is at 0 eV and
difference curve shows distinguishing features
for each of the DNA bases.} }
\label{f3}
\end{minipage}
\end{figure*}

\section{Results and Discussion}

\begin{figure*}[htpb]
    \begin{minipage}[!t]{0.70\linewidth}
   \epsfig{file=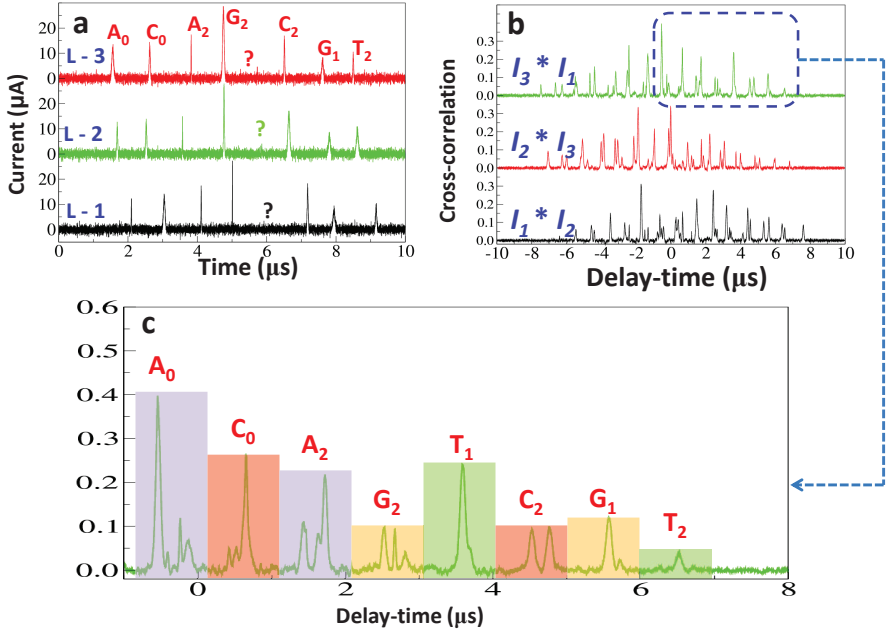, width=\linewidth}
    \end{minipage}\hfill
    \begin{minipage}[!t]{0.28\linewidth}
\caption{\small{ {\bf {a}} Current vs. time ($\mu$s) plot for a translocating DNA sequence `ACAGTCGT'
for three
graphene layers labeled as L-1, L2, and L-3. An additive white noise is
included in the current spectrum. Due to high noise to signal ratio some of
the spectral features became harder to recognize (indicated by a question mark in the figure).
{\bf {b}}
Cross-correlation between current signals {\it {I$_1$(t)}}, {\it {I$_2$(t)}}, and
{\it{I$_3$(t)}} as functions of delay time $\Delta t$ , where the currents
are from graphene layers L-1, L-2, and L-3 respectively.
{\bf {c}}
Enlarged segment of the cross-correlation function from (b). These
correlation-signal peaks correspond to the peaks from current-signal for
the DNA sequence ACAGTCGT.}}
\label{f4}
\end{minipage}
\end{figure*}

We first discuss our first-principles calculations of transmittance for individual DNA bases inside
the graphene nanopore, as presented in ~\ref{f3}. Then in ~\ref{f4}, we show the partial signal recovery using our
time-simulation model with three layer graphene nanopores and the cross-correlation between the corresponding signals.

In our first-principles approach,
for each DNA base, we have taken three random angular orientation with the graphene membrane, while
calculating the transmittance between the two electrodes with 0.7 V bias voltage. The configuration averaged
transmittance for
A, C, G, and T are shown in the solid blue curve in ~\ref{f3}(a)-(d). The conductance of a pure graphene nanoribbon
with hydrogenated nanopore is shown
in solid red curve in ~\ref{f3} for comparison. The transmittance curve is analogous to the
non-equilibrium density of states in the
presence of the bias voltage where the zero of energy is the Fermi energy of the central graphene region.
The vertical dashed lines are at -0.35 eV and +0.35 eV, which are the
chemical potentials of the left and right electrodes respectively. For each base (~\ref{f3}(a)-(d)),
the transmittance curve (solid blue line)
in between
the left and right electrode chemical potentials is significantly enhanced compared to the pure graphene membrane with a nanopore (solid red
line). The features in this region are characteristic of the four bases. For example, a comparison of the
Guanine transmittance
(~\ref{f3}(c)) with that of Thymine (~\ref{f3}(d)), shows the presence of a characteristic broad peak.

For a systematically study of the difference between the transmittance among the four bases, we also plotted the
difference curves (the top three) in ~\ref{f3}(a)-(d). If the signatures  of one or more of the DNA bases are known prior to the detection,
the difference curve may provide the signature of an unknown base. For example, if one knows the
transmittance of Thymine, a comparison of the characteristic features of difference-squared transmittance
$(A-T)^2$, $(G-T)^2$, $(C-T)^2$, helps identify the unknown base.
~\ref{f3}(a),(c), and (d) show the difference-curves contain several (up to three) dominant peaks in between the vertical dashed
lines. In principle, it is possible to calculate a
large number
of configurations and maintain a complete data-base of such characteristic difference curves for the
sequencing purpose.

Such methods are challenged by two major limitations. The first one is
prior knowledge of the
exact location of one or more kinds of DNA base, either from the transmittance curve or form other technique.
The second one
is the presence of significant noise in the data, which makes it difficult for the detection of any single base.
Some bases exhibit characteristic
features in the
transmittance curve, which make them easily detectable. For example, the Thymine
(~\ref{f3}(d) solid blue line) has a very low conductance compared to the others which (in agreement with previous
calculations~\cite{ventra_2005,ventra_2006}) shown by the low peak
amplitude near 0 eV. However, even the detection of Thymine can be difficult in the presence of noisy data.
To illustrate the specifics of the approach, we present the simulation of a time-series for three graphene
nanopore layers with the test
sequence
A$_0$C$_0$A$_2$G$_2$T$_1$C$_2$G$_1$T$_2$ in Fig.4.

In nanopore based DNA sequencing, the  current ($I(t)$) is the measured quantity rather than the
transmittance ($T(E)$). Thus, we calculated the current from the transmittance.
Using the
parameters described previously we simulated time-dependent current spectra $I_{L-1}$, $I_{L-2}$, and
$I_{L-3}$ for our test sequence, as shown in ~\ref{f4}(a).
The low current amplitude for Thymine in the case of T$_1$ and
T$_2$ is expected from the transmittance curve in ~\ref{f3}(d), but the natural noise present in the data makes it difficult
to confirm the presence of T$_1$ at the expected location. In ~\ref{f4}(b), we present the cross-correlation between the
current spectra from different pairs of graphene layers.
For each pair, the cross-correlation is plotted as a function of
time-delay within the -10 $\mu$s to +10 $\mu$s range. The cross-correlation
spectrum is approximately
symmetric around mid point of the total range due to the overlaps between
similar
pairs of peaks from opposite ends of the original data. Therefore, we
only focus on the positive time-delay. The correlation spectrum inside the
highlighted dashed box in ~\ref{f4}(b) is enhanced
in ~\ref{f4}(c). By comparing peaks between ~\ref{f4}(a) and (c), we confirm the presence of Thymine with T$_1$ configuration. Although the amplitudes of the current spectrum do not translate directly into the
amplitudes of the
cross-correlation spectrum, they confirm the existence of T$_1$.
Thus,
a time-series
analysis using current cross-correlations
$\langle I_{i}(t) \otimes I_{j}(t) \rangle$
recovers all eight peaks in our test sequence (~\ref{f4}(b)).
The suppression of
white noise is substantial and the peaks at time-delay=0 in the
correlation function (~\ref{f4}(b)) are enhanced.

\begin{figure*}[htpb]
    \begin{minipage}[!t]{0.70\linewidth}
   \epsfig{file=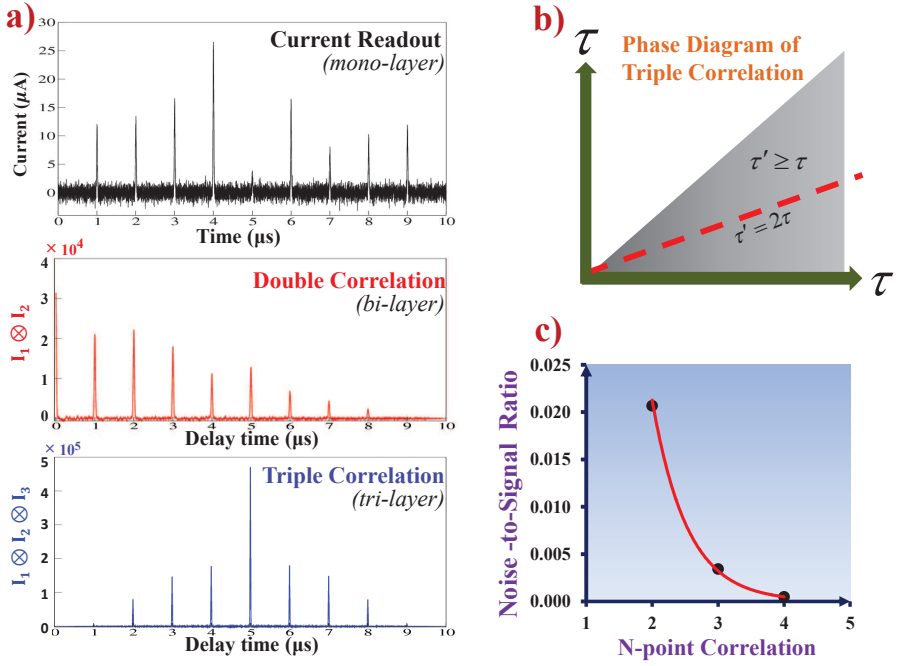, width=\linewidth}
    \end{minipage}\hfill
    \begin{minipage}[!t]{0.28\linewidth}
\caption{\small {The three
panels in (a) show the improvements in signal-to-noise ratio with higher order cross-correlation.
Time dependent current spectrum for the sequence
A$_0$C$_0$A$_2$G$_2$T$_1$C$_2$G$_1$T$_2$C$_1$ from a single layer graphene is shown in the top
panel (black);
where the double and triple cross-correlated spectrums are shown in the middle (red) and bottom (blue)
panels. (b) Phase diagram of a triple correlation function on a 2D delay-time parametric space
for $\tau$ and
$\tau'$. The dashed red line is our constrain for calculating the triple correlation function.
(c) Nearly exponential decay of noise-to-signal ratio with higher order correlation.}}
\label{f5}
\end{minipage} 
\end{figure*}

We can  easily extend this approach to
three-point or higher $N$-point correlations, which
we demonstrate here, to exponentially reduce the noise-to-signal ratio.
The two-point cross-correlation is generally expressed with a single parameter as in
\begin{equation}
R^{(2)}(\tau)=\int_0^T I_1(t) I_2(t-\tau) dt,
\end{equation}
where the time interval is between $0$ to $T$. The three-point correlation is a function
of two independent variables
\begin{equation}
R^{(3)}(\tau, \tau')=\int_0^T I_1(t) I_2(t-\tau) I_3(t-\tau') dt.
\end{equation}
We can simplify the description of triple correlation function in the complete two dimensional
parametric space by constraining it to the line $\tau' = 2 \tau$ as in ~\ref{f5}(b).
Thus the constrained triple correlation function becomes,
\begin{equation}
R^{(3)}(\tau)=\int_0^T I_1(t) I_2(t-\tau) I_3(t-2 \tau) dt.
\end{equation}
Following this procedure we can measure currents from $N$ independent graphene layers and
calculate constrained $N$-point correlation as
\begin{align}
R^{(N)}(\tau)=\int_0^T I_1(t)&I_2(t-\tau) I_3(t-2 \tau) .... \nonumber \\
&.... I_N(t-(N-1)\tau) dt.
\end{align}
The three panels in ~\ref{f5}(a) show our calculated current signal from a single
layer as well as the
two and
three point cross-correlation functions from the corresponding two and three independent graphene
nanopores. The test sequence used here is
A$_0$C$_0$A$_2$G$_2$T$_1$C$_2$G$_1$T$_2$C$_1$.
Using two, three and four point cross-correlation functions, we estimated the ratios
between
the average
signal and average noise in each case, as shown in Table.1 in the supplementary section. We confirm
the exponential drop in the noise to signal ratio
as shown in ~\ref{f5}(c). The computational details and the table
containing the results are also given in the supplementary section.

\section{Computational Method}

In this work, we ignore the background contribution from the large phosphate backbone typically present in a single stranded DNA (ssDNA). This
simplification is based on the assumption
that by identifying and
subtracting the background noise coming from the heavy and rigid backbone structure.
one can isolate the relevant signal from the individual bases. More specifically we have built
on earlier work~\cite{towfiq_dna,Kilina2011,ventra_2005,scheicher} to model the pore conductance
containing a molecule in two steps:
1) First,
we
carried out {\it {ab initio}} calculations of transmission ($T(E)$) and current ($I$) as a single
DNA base translocates through the nanopore of a graphene mono-layer.
2) Then, we simulate the time-dependence of the current data by adopting a simple model with multi-layered graphene nanopores
with added statistical noise and broadening.

Calculations of transmission were performed taking each DNA base inside the nanopore with three
different angular orientations, and using the Landauer-Buttikker~\cite{land} formalism implemented
in the {\it ab initio} software ATK~\cite{mads_2}. We emphasize that out approach does not rely nor requires
a geometry optimization of molecules in the pores. The translocation is a dynamical process with significant
variations of configurations found for molecules inside a pore. Thus, the same
molecule can arrive in different orientations at each pore, a process which contributes to the configuration noise sources
that we address here.
Therefore, we do not optimize the configurations and instead use the set
of various configurations as the set, from which the random sampling is taken.

 In these calculations, we have taken a graphene nanoribbon with
208 carbon atoms in the conduction region, where the nanopore is constructed by removing center carbon atoms and capping the inner wall
with hydrogen atoms, since hydrogenated edges were found~\cite{scheicher} to enhance the average experimental conductivity.
The bias voltage between the left and right electrodes
is fixed as +0.35 and -0.35 eV.
In this work, the nanopore dimension is much smaller than that modeled
by other groups~\cite{postma,prezhdo}. The details and various parameters of our first-principles calculations can be found in the supplementary section.

To demonstrate the recoverability of
current ($I(t)$) signals from noise, we show the relation between
noise coming from different layers. For simplicity, we consider the
dominant noise primarily from two sources. As the bases translocate
through the
{\it {i}}-th graphene nanopore layer, the vibration in the DNA backbone
may influence individual base plane to land with random angular orientation
with the
graphene plane, causing a configuration-noise $S_i^C(t)$.
The additional noise, such as
thermal vibration of the graphene membrane at the $i$-th nanopore,
is defined as $S_i^A(t)$. Thus the total noise
of $i$-th nanopore can be expressed as
\begin{equation}
S_i(t)=S_i^C(t)+S_i^A(t).
\end{equation}
The correlation between the two layers is therefore given by
\begin{align}
<S_i(t)&\cdot S_j(t')>\,=\,<S_i^C(t) \cdot S_j^C(t')> \nonumber \\
&+<S_i^C(t) \cdot S_j^A(t')>
+<S_i^A(t) \cdot S_j^C(t')> \nonumber \\
&+<S_i^A(t) \cdot S_j^A(t')>.
\end{align}
Here $t'=t+\Delta t$.
For $i \neq j$, the contribution from the last three terms on the right side of
Eq.~2  are negligible due the weakly or uncorrelated signals in separate nanopores. Since the
DNA bases are strongly attached to the ssDNA backbone, the configuration-noise
between two membranes mainly contributes to the
first term in Eq.~2. Therefore, the noise can be approximated as
\begin{equation}
<S_i \cdot S_j>\,\approx\,<S_i^C \cdot S_j^C> ,
\end{equation}
where, for $i = j$, all the terms on right side of Eq.~2 survive and
contribute significantly to the total noise. Since the noise between $i$ and $j$ is uncorrelated, a comparison
of their signals will enhance the individual base signals by reducing the noise to signal ratio.

There are two extreme limits in which we can take advantage of the above observation. These limits relate
to the rate of base translocation compared to the typical vibrational frequency of the
bases facing the electrodes. When this occurs,
the above cross correlations allow us to reduce the {\it intrinsic} noise due to random
orientations. On the other hand, when the translocation rate is slower than the vibrational frequency,
the uncorrelated noise is eliminated and the only one that survives is the correlated one.
We focus here on the second case since experimentally the latter situation is
more likely~\cite{Zwolak2008,Branton2008}.

As an example, we show the low current amplitude for Thymine in ~\ref{f4}(a), and in ~\ref{f4}(c) the
enhancement of the signal to noise ratio.
We have taken a
test sequence
A$_0$C$_0$A$_2$G$_2$T$_1$C$_2$G$_1$T$_2$,
where the subscripts
imply different angular orientations of the
bases inside the pore. The time dependence of this sequence is modeled by taking the time
interval between two consecutive bases $ \tau = 1.0 \;\mu\text{s}$,
including a random Gaussian
uncertainly between the interval with $\sigma_{\tau}= \pm 0.2 \;\mu\text{s}$. Each current signal is
also broadened using a random Gaussian broadening with $\sigma_{broad}=0.2\,\mu A$. To simulate a realistic experiment
with background noise, we have also included additive white
Gaussian noise.
We assume that with the applied field in the vertical direction,
the average elapsed time between two translocating bases is
$\tau \approx 1.0 \;\mu\text{s}$. The time-distance between two consecutive graphene layers is set to  $\Delta t
\approx 0.2 \;\mu\text{s}$.

\section{Conclusions}

We implement first-principles calculation of transmittance for a systematic study of the identification of single DNA bases or other biomolecules
translocating through graphene nanopores.
To eliminate the high background noise, we propose a multilayered
graphene-based nanopore device combined with a multi-point cross-correlation method to substantially
improve
the signal to noise ratio of the electronic readout of biomolecules.
To illustrate this approach, we adopted a statistical method for
simulating the time-dependent current spectrum. The enhanced resolution is produced by the multiple translocation readouts of the same bases of the same molecule through the pores.
The cross-correlated signals from each pair of electrodes will suppress the uncorrelated noise
produced by each single translocation event.

In this way, thymine can serve as a ``reference molecule'' for identifying other molecules from
the difference transmittance curves. We also demonstrate the recovery of signals associated with different configurations
by taking cross-correlations between different pairs of graphene layers. This study provides a promising
method for an enhanced signal to noise ratio in the multipore graphene based devices (or any other serial sequencing device), and
their potential applicability as a next generation biomolecular detection technique. While we focus on the correlations in DNA bases, this cross-correlation method can be used for any molecule or molecular series for detection or identification purposes.

\acknowledgement
We are grateful to K.T. Wikfeld, K. Zakharchenko  and Svetlana Kilina for
useful discussions. This work is supported by the Center for Integrated
Nanotechnologies at Los Alamos, a U.S. Department of Energy, Office of Basic
Energy Sciences user facility. Los Alamos National Laboratory, an
affirmative action equal opportunity employer, is operated by Los Alamos
National Security, LLC, for the National Nuclear Security Administration
of the U.S. Department of Energy under contract DE-AC52-06NA25396. Work at
NORDITA was supported by VR 621-2012-2983 and ERC 321031-DM.
MD acknowledges partial support from the National Institutes of Health.
IKS is supported by AFOSR FA9550-10-1-0409.


\providecommand*\mcitethebibliography{\thebibliography}
\csname @ifundefined\endcsname{endmcitethebibliography}
  {\let\endmcitethebibliography\endthebibliography}{}

\end{document}